\documentclass{article}
\usepackage{spconf,amsmath,epsfig}
\usepackage{xcolor,soul,multirow}
\sethlcolor{yellow}
\usepackage{caption}
\usepackage{subcaption}
\usepackage{booktabs}
\usepackage{amsfonts}
\usepackage{hyperref}
\let\OLDthebibliography\thebibliography
\renewcommand\thebibliography[1]{
  \OLDthebibliography{#1}
  \setlength{\parskip}{0pt}
  \setlength{\itemsep}{0pt plus 0.3ex}
}

\begin{document}\sloppy

\def\x{{\mathbf x}}
\def\L{{\cal L}}
\newcommand\average[1]{\left\langle#1\right\rangle}
\def\method{\textbf{Method name placeholder}}

\title{Learn how to prune pixels for multi-view neural image-based synthesis}

\name{Marta Milovanović$^{\star}$ \qquad Enzo Tartaglione$^{\star}$ \qquad Marco Cagnazzo$^{\star \ddagger}$ \qquad Félix Henry$^{\mathsection}$}

\address{
      $^{\star}$ LTCI, Télécom Paris, Institut Polytechnique de Paris, France \\
      $^{\ddagger}$ DEI, University of Padua, Italy \\
      $^{\mathsection}$ Orange Labs, France}

\maketitle

\begin{abstract}
Image-based rendering techniques stand at the core of an immersive experience for the user, as they generate novel views given a set of multiple input images. Since they have shown good performance in terms of objective and subjective quality, the research community devotes great effort to their improvement. However, the large volume of data necessary to render at the receiver's side hinders applications in limited bandwidth environments or prevents their employment in real-time applications. We present LeHoPP, a method for input pixel pruning, where we examine the importance of each input pixel concerning the rendered view, and we avoid the use of irrelevant pixels. Even without retraining the image-based rendering network, our approach shows a good trade-off between synthesis quality and pixel rate. When tested in the general neural rendering framework, compared to other pruning baselines, LeHoPP gains between $0.9$ dB and $3.6$ dB on average.
\end{abstract}
\section{Introduction}
\label{sec:intro}
Immersive video technologies enable viewers to perceive a scene as if they are inside the video. However, they still pose a challenge to the industry: to provide a novel viewpoint to the viewer, a high amount of data needs to be transmitted and processed. This poses constraints that include power consumption, transmission bandwidth, and hardware capabilities.

Lately, there has been huge progress in the field of neural image-based rendering~\cite{Penner2017Soft3R,mildenhall2019llff}. These methods aim to render novel views of complex scenes captured by a sparse set of input views, and they can have different 3D scene representation formats. The field of neural image-based rendering has been transformed by the Neural Radiance Fields (NeRF) technique~\cite{nerf}. By utilizing multi-layer perceptrons (MLP) and positional encoding, it can represent the continuous 5D radiance field with remarkable accuracy. The model's ability to handle non-Lambertian effects while rendering new views is enabled by its view-dependent nature. Nevertheless, NeRF-based models encode a scene into their own weights and require a lengthy optimization process to render an unfamiliar scene, which makes them impractical. 

One of the recently developed methods, called IBRNet~\cite{ibrnet}, is a state-of-the-art, robust neural radiance field method. It generalizes well to novel scenes, owing to its end-to-end design that prioritizes the optimization of synthesis quality. Moreover, it outperforms recent one-shot synthesis approaches, like~\cite{mildenhall2019llff}, while still having good quality synthesis results as compared to single-scene inference approaches. Furthermore, the current technological progress allows us to have a neural (image-based) renderer on the client side, which produces high-quality synthesis results but requires many available source views of a given scene. This creates a big obstacle to its current deployment using existing standards (such as MPEG Immersive Video - MIV~\cite{mivfdis}, a standard for volumetric video transmission). 

In this paper, we leverage the generalization capabilities of IBRNet for a use-case scenario of immersive video processing and potential data transmission, by taking into consideration the need for a pixel rate reduction. Pixel rate is defined as the number of luma pixels per second that a compliant decoder should be able to process. Constraints for pixel rate are defined in the MIV Common Test Conditions (CTC)~\cite{CTC}. Thus, we propose a method that distinguishes important pixels from the ones that can be pruned. For each input view, our method creates a corresponding pruning mask and removes the non-essential pixels. Pruning mask computation is guided by the loss function computed on target views rendered with IBRNet, with respect to the input views. Then, in the final stage, a target view is rendered using the pruned source views. 

Pixel rate is a very important constraint for the feasibility of a multi-view video transmission system, therefore, our work is a significant first stage in the full immersive coding pipeline. To the best of our knowledge, this is the first paper that studies the impact of pixel pruning on the multi-view, texture-based neural renderer, more specifically, IBRNet. Hence, we also propose a study with random pruning baselines, and we compare them.
We observe the better performance of our approach against randomly pruned pixels, and an overall good compromise between rendered view quality and pixel rate, even without retraining and extra fine-tuning per video sequence for IBRNet.\footnote{we used the pre-trained model available at: \url{https://github.com/googleinterns/IBRNet}}
The rest of the paper is organized as follows. Section~\ref{background} gives an overview of the IBRNet architecture, volumetric video standard, and related pixel pruning techniques. Section~\ref{proposal} describes the proposed approach. Section~\ref{results} presents experimental conditions, results, and analysis, whereas Section~\ref{conclusion} concludes the paper.

\section{Background}
\label{background}
\subsection{IBRNet}
\label{sec:ibrnet}
IBRNet~\cite{ibrnet} is a new technique that builds both on traditional image-based rendering and on the NeRF approach~\cite{nerf}, generating a continuous scene radiance field. It has better performance when rendering challenging scenes than previous IBR methods because it is designed in an end-to-end manner to optimize for synthesis quality. Moreover, in contrast to NeRF, it generalizes to arbitrary new scenes, which is a desired characteristic for an immersive video renderer in the transmission scenario. IBRNet first identifies a set of neighboring source views to a target and it extracts their image features. Then, for each camera ray in the target view, it computes colors and densities for a set of samples along the casted ray. Finally, colors and densities along that ray are accumulated to render the color. In the course of training, the mean squared error between the ground truth pixel color and the rendered pixel color is minimized.

The IBRNet architecture is as follows. First, the image features extracted from all source views are given to an MLP architecture, to aggregate local and global information and obtain features that are multi-view aware. Subsequently, a ray transformer module is used to aggregate information of all samples computed on the casted ray, which improves geometric reasoning and density predictions.

\subsection{Volumetric video standard}
The rising consumer interest in immersive technologies has triggered the activity in the MPEG standardization community~\cite{wien_standardization_2019}. Recently, MPEG released a new specification to facilitate the storage and transmission of digital immersive media for playback with 6 degrees of freedom, called MPEG Immersive Video (MIV)~\cite{mivfdis,9897142}. The MIV main profile supports multi-view plus depth input data format, where each (texture) viewpoint has a corresponding depth map that represents the distance of the objects in the scene to the camera. The MIV reference software (TMIV)~\cite{TMIV} consists of a normative decoding process (based on MIV metadata) and a non-normative rendering process, where a targeted view is generated using depth image-based rendering (DIBR) methods~\cite{RVS}. 

MIV is designed to leverage already developed end-to-end video coding systems; however, coding depth maps using legacy video codecs is not an easy task, because of the nature of depth data. Since depth poses an additional burden in terms of bitrate and pixel rate, many approaches have been developed to alleviate this overhead, one of which is decoder-side depth estimation. This encouraged a new profile of MIV, called Geometry Absent (GA) profile~\cite{dm-dsde}, which transmits only the texture information in the bitstream. Still, this method uses depth estimation tools to compute the depth maps before rendering the target view with DIBR methods.

\subsection{Pixel pruning}
\label{sec:pruning}

 Input pixel pruning has already been pioneered in the context of scalable self-supervised learning in computer vision, where random image patches are masked in the input image and the missing pixels are subsequently reconstructed~\cite{he2022masked}. Moreover, a recent approach proposes an online selection of context points for efficient meta-learning, on 2D videos and other data modalities~\cite{tack2023efficient}. Yet, these methods assume that the context (input image) is the same as the target, as opposed to scene rendering, where inputs are different from the target.

Regarding multi-view video pixel pruning, TMIV offers a depth-based solution. Pruning in TMIV~\cite{TMIV} is motivated by defined hard limits on the pixel rate~\cite{pixel-rate}, based on hardware limitations, and it is a method for removing redundant samples among the input views in TMIV. Before pruning, the input views are labeled as basic (to be sent as complete) or additional (to be pruned and transmitted only in portions). Subsequently, pruning determines which pixels in additional views are already present either in basic views or other additional views and therefore not needed for transmission and rendering. On the contrary, pixels that cannot be recovered from other views, are preserved. Then, a pixel is pruned if the difference between the synthesized and source depth/texture luma value does not exceed a given threshold. A second-pass pruning~\cite{shin2022enhanced} based on global color matching is employed to restore some of the pixels, that were initially pruned due to depth imperfections or illumination changes among different source textures.

\section{Proposed method}
\label{proposal}

\begin{figure}
     \centering
     \begin{subfigure}[b]{\columnwidth}
         \centering
         \includegraphics[width=\textwidth]{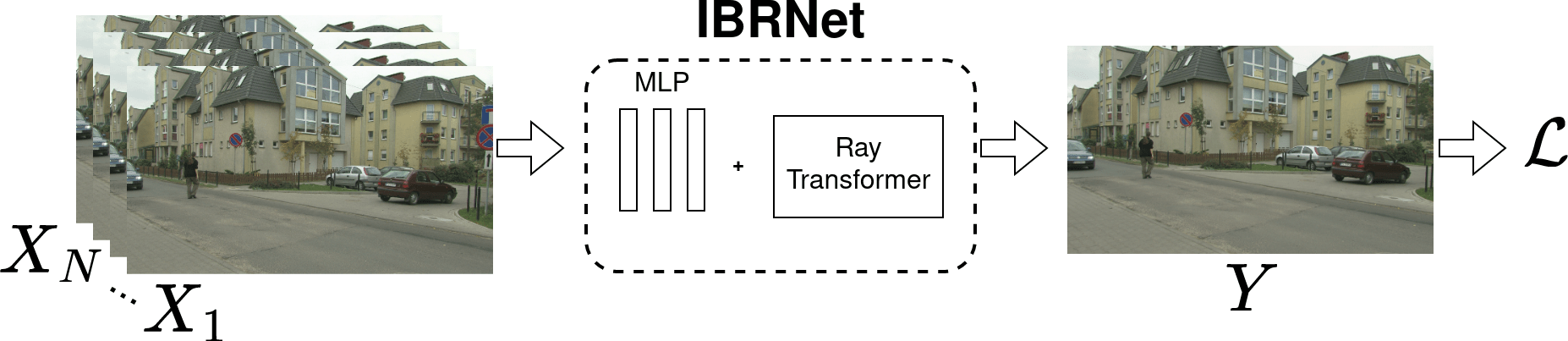}
         \caption{Rendering loss computation in a forward pass.}
         \label{fig:methodpart1}
     \end{subfigure}
     \begin{subfigure}[b]{\columnwidth}
         \centering
         \includegraphics[width=\textwidth]{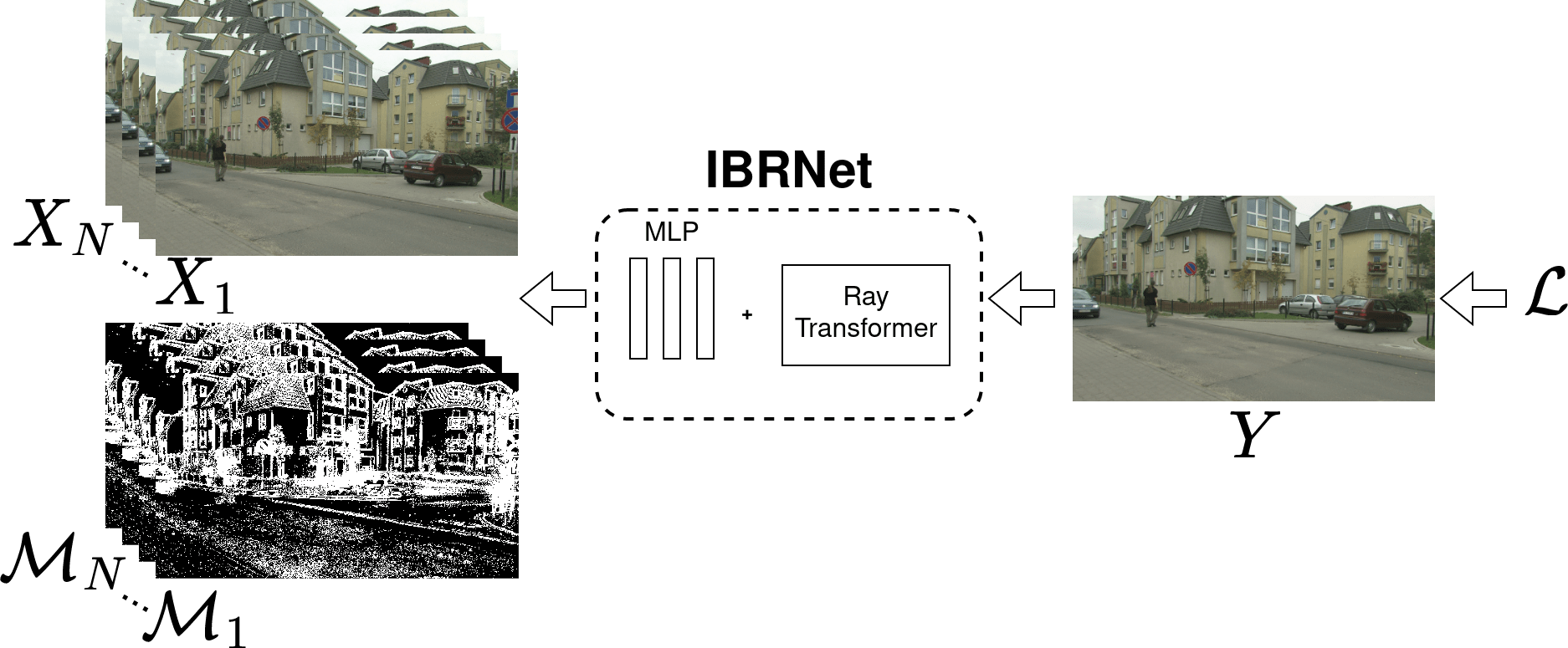}
         \caption{Computation of importance and mask in a backward pass.}
         \label{fig:methodpart2}
     \end{subfigure}
     \begin{subfigure}[b]{0.88\columnwidth}
         \centering
         \includegraphics[width=\textwidth]{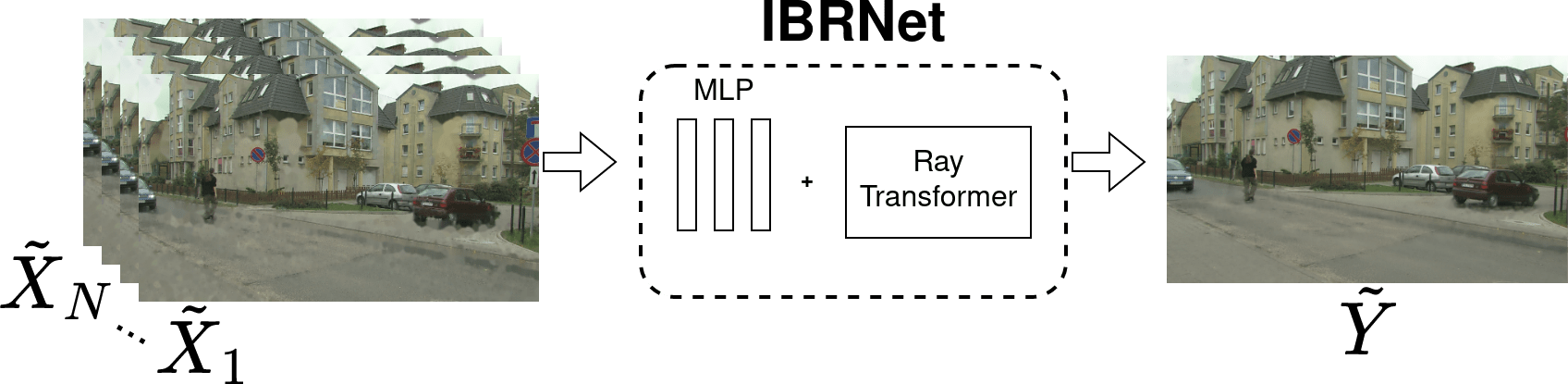}
         \caption{Rendering result, from pruned and inpainted images.}
         \label{fig:methodpart3}
     \end{subfigure}
        \caption{Overview on the LeHoPP method application.}
        \label{fig:method}
\end{figure}

This section describes our proposal, called LeHoPP, which is a pruning approach performed at the pixel level of input images, in the image-based rendering setup. Fig.~\ref{fig:method} provides a scheme of the method. In this section, we will first present how to perform the computation for the pruning mask, and then we will provide an algorithmic overview of LeHoPP.

\subsection{Pruning mask computation}
\label{sec:mask}
Let us consider the multi-view input images $X_i\in \{0,1,...,255\}^{W_i\times H_i\times 3}$ with $i=\{1,...,N\}$, where $N$ is the number of source views, and $W_i$ and $H_i$ are the width and the height for the $i$-th view. Before feeding $X_i$ to IBRNet, the images are pre-processed, with standardization and scaling, which transforms $X_i$ to $\hat{X}_i\in \mathbb{R}^{W_i\times H_i\times 3}$, according to the standard implementation of IBRNet~\cite{ibrnet}. We will associate to every $X_i$ and $\hat{X}_i$ a pruning mask $\mathcal{M}_i\in \{0,1\}^{W_i\times H_i}$. This mask indicates whether some pixels should be either preserved $(1)$ or discarded $(0)$. 

To obtain the values of each image mask, we first need to compute an importance score $\mathcal{I}_i$, associated with every pixel of the source view. The importance function tells us how much are rendered images sensitive to changes in a certain pixel of the input image. We know that given the loss function $\mathcal{L}$ we optimize our masks on (mean squared error on the target view, coherently with the loss employed for training IBRNet), we can estimate some variations in the value for the loss function by Taylor series expansion as
\begin{equation}
    \Delta \mathcal{L} \approx \frac{\partial \mathcal{L}}{\partial \hat{X}_{i}(u,v)}\Delta \hat{X}_{i}(u,v).
    \label{eq:taylor}
\end{equation}
From this, we can define our importance score as 
\begin{equation}
\label{eqn:importance}
    \mathcal{I}_{i}(u,v) = \bigg |\dfrac{\partial \mathcal{L}}{\partial \hat{X}_{i}(u,v)}\bigg | \cdot \left|\hat{X}_{i}(u,v)-\hat{X}_{i}^{\text{inp}}(u,v)\right|,
\end{equation}
where $u$ and $v$ denote pixel coordinates in an image, and $\hat{X}_{i}^{\text{inp}}$ denotes an inpainted approximation of $\hat{X}_{i}$. To maximize the performance of the pruning mask $\mathcal{M}_i$, the source importance score $\mathcal{I}_i$ is cumulated over image $(R,G,B)$ channels and averaged over all the target views. Since different subsets of nearby source views are used to render the target views, we average the gradient values that we get for a particular source view but from different target loss computations.

Unlike similar approaches in the literature~\cite{tartaglione2018learning, Deng_2023_WACV}, given that masked values are discarded pixels, we can run an inpainting algorithm to partially recover lost information. Hence, our $\Delta \hat{X}_{i}(u,v)$ is not simply the value of the pixel (as traditionally done in any pruning algorithm), but it is the distance between the real value and the estimation of its inpainted value. For computational efficiency, we estimate it by averaging the value of neighboring pixels:
\begin{equation}
    \hat{X}_{i}^{\text{inp}}(u,v) = \frac{1}{8} \left[-\hat{X}_{i}(u,v)+\sum_{i=u-1}^{u+1}\sum_{j=v-1}^{v+1} \hat{X}_{i}(i,j)\right]\!.
    \label{eq:inp}
\end{equation}
Here, inpainting is not used to fill out the occluded regions in the images after the novel view rendering, rather it is used after the pixel pruning because IBRNet is trained on full images. Instead of giving the black/gray pruned pixels to the IBRNet renderer, we are inpainting the pruned regions and incorporating their impact in our importance score.

We use the same mask for all the frames of the intra-period, to reduce the frequency of mask updating and overhead for potential transmission. Therefore, the importance score is cumulated over all the frames of one intra-period ($16$ frames).  Finally, given that we target the remotion of the $\gamma\in [0;1]$ fraction of pixels from the input view, we can use the quantile function $\mathcal{Q}_\mathcal{I}(\gamma)$ to determine the threshold to apply to importance values, to finally compute the pruning mask:
\begin{equation}
    \label{eq:mask}
  \mathcal{M}_i(u,v)=\begin{cases}
    1 & \text{if $\mathcal{I}_{i}(u,v,c) \geq \mathcal{Q}_\mathcal{I}(\gamma)$,}\\
    0 & \text{otherwise}.
  \end{cases}
\end{equation} 

\subsection{Overview of LeHoPP pipeline}
The scheme of LeHoPP is depicted in Fig.~\ref{fig:method}, and it consists of three steps. First, the multi-view images $X_i$ are given to IBRNet and a forward propagation step is performed, and the loss $\mathcal{L}$ is computed from the synthesized view $Y$ (Fig.~\ref{fig:methodpart1}). Then, through back-propagation, we can calculate $\frac{\partial \mathcal{L}}{\partial \hat{X}_i}$, from these the importance scores $\mathcal{I}_i$ according to \eqref{eqn:importance} and the masks are computed according to \eqref{eq:mask} (Fig.~\ref{fig:methodpart2}). Finally, the masks are applied on the input images to prune the pixels, which are then inpainted, and the resulting images $\tilde{X}_i$ are given to the IBRNet renderer, to synthesize the target view $\tilde{Y}$ (Fig.~\ref{fig:methodpart3}).

\section{Results and discussion}
\label{results}

\subsection{Test conditions}
Our dataset consists of eight perspective sequences from MIV CTC~\cite{CTC}, with natural and computer-generated multi-view content of high resolution, captured by a sparse camera setup. We compare LeHoPP with two baselines: \emph{base1} ($32\!\times\!32$ block-based random pixel pruning) and \emph{base2} ($4\!\times\!4$ block-based random pixel pruning). Since we did not find any existing method for addressing the problem of pixel pruning as in our context, we decided to compare the proposed method with a simple, ``blind'' approach (random pruning). We remark that the TMIV pruner is not a relevant comparison in our setup, since it is designed for a different goal. It is based on depth image-based rendering - while IBRNet deals with texture-only data - and yields a small set of full views together with small image areas called patches, which cannot be used in the IBRNet rendering context. Moreover, the texture videos pruned with our method cannot be utilized in the TMIV rendering, because we do not rely on depth maps.

In our experiments, the pruning percentage $\gamma$ is the only varying parameter, demonstrating the trade-off between quality and pixel rate. Furthermore, pruned pixels are inpainted prior to being input to the rendering, using Telea~\cite{telea} method from the OpenCV library. The number of input source views for rendering is set to 9, for all the sequences. We evaluate the quality of synthesized views using PSNR, SSIM~\cite{ssim}, and LPIPS~\cite{lpips} metrics, as proposed for IBRNet.

\begin{table*}[t]
\small
\centering
\caption{Results for different pixel pruning percentages: $5\%$, $10\%$ and $20\%$.}
\label{tab:results}
\begin{tabular}{c c c c c c c c c c c c} 
 \toprule
 \multirow{2}{*}{\bf Metric}&\multirow{2}{*}{$\boldsymbol{\gamma}$}&\multirow{2}{*}{\bf Method}& \multicolumn{9}{c}{\bf Sequences}\\
 &&&            Painter &Frog   &Carpark    &Fan    &Shaman &Hall   &Street &   Mirror& \bf Average\\
 \midrule
 \multirow{10}{*}{PSNR ($\uparrow$)}
 &0\% &Anchor      &31.14      &28.18  &38.96       &37.67  &47.06  &39.85  &43.36  &33.38  &37.45\\
 \cline{2-12}
 &      &\emph{base1}  &29.89     &27.25  &35.79      &33.62  & 41.41 &38.11  &37.78  & 30.43 & 34.29\\  
 &5\%   &\emph{base2}  &30.72     &27.75  &37.40      &35.17  & 43.42 &39.23  &40.49  & 31.64 & 35.73\\
 &      &LeHoPP &\textbf{31.04}     &\textbf{27.98}  &\textbf{38.76}      &\textbf{37.40}  &\textbf{44.76}  & \textbf{39.78} &\textbf{42.73}  & \textbf{33.13} & \textbf{36.95}\\
 \cline{2-12}
 &      &\emph{base1}  &29.03     &26.43  &33.62      &31.38  &38.58 &36.78  &35.01  & 29.08 &32.49\\  
 &10\%   &\emph{base2}  &30.31     &\textbf{27.30}  &36.05     &33.57  &41.29 &38.65  &38.58  & 30.57 & 34.54\\
 &      &LeHoPP &\textbf{30.44}     &27.27  &\textbf{38.45}      &\textbf{36.76}   &\textbf{41.74}  &\textbf{39.69}  &\textbf{42.10}  &\textbf{32.37} &\textbf{36.10}\\
 \cline{2-12}
 &      &\emph{base1}  &27.74     &24.96       &30.31      &28.57      &35.36     &34.80   &31.57  &27.22 &30.07\\  
 &20\%   &\emph{base2}  &\textbf{29.48}     &\textbf{26.40}      &33.78      &31.21     &\textbf{38.46}      &37.49   &35.78  &29.05  & 32.71 \\
 &      &LeHoPP & 28.04 &   24.90       &\textbf{37.39}      &\textbf{34.26}     & 34.55     &39.42    &\textbf{40.37}  &\textbf{29.75}& \textbf{33.58}\\
  \midrule
 \multirow{10}{*}{SSIM ($\uparrow$)}
 &0\%&Anchor      &0.9561     &0.9092  &0.9901     &0.9926    &0.9982 &0.9965 &0.9953    &0.9684 &0.9758\\
 \cline{2-12}
 &      &\emph{base1}  &0.9450     &0.8941  &0.9788      &0.9784  & 0.9911 &0.9934  &0.9853  & 0.9538 & 0.9651\\  
 &5\%   &\emph{base2}  &0.9484     &0.8962  &0.9819     &0.9819  & 0.9930 &0.9948  &0.9881  & 0.9575 & 0.9677\\
 &      &LeHoPP &\textbf{0.9511}     &\textbf{0.9038}  &\textbf{0.9878}      &\textbf{0.9913}   &\textbf{0.9952}  & \textbf{0.9963}  &\textbf{0.9930}  & \textbf{0.9675} & \textbf{0.9733}\\
 \cline{2-12}
 &      &\emph{base1}  &0.9337     &0.8771  &0.9652      &0.9626      &0.9827  &0.9914 &0.9739  &0.9388 & 0.9532\\  
 &10\%   &\emph{base2}  &0.9397     &0.8820 &0.9718      &0.9695     &0.9869  &0.9929 &0.9797  &0.9455  & 0.9585 \\
 &      &LeHoPP &\textbf{0.9412}      &\textbf{0.8923} &\textbf{0.9844}      &\textbf{0.9872}     &\textbf{0.9871}   &\textbf{0.9961}  &\textbf{0.9894}  &\textbf{0.9640} & \textbf{0.9677}\\
 \cline{2-12}
 &      &\emph{base1}  &0.9103     &0.8390      &0.9333      &0.9272      &0.9641     &0.9855 &0.9473  &0.9069 & 0.9267\\  
 &20\%   &\emph{base2}  &\textbf{0.9197}    &0.8505      &0.9468      &0.9403     &\textbf{0.9722}      &0.9883 &0.9594  &0.9180  & 0.9369\\
 &      &LeHoPP &0.9061     &\textbf{0.8509}      &\textbf{0.9745}      &\textbf{0.9657}     &0.9473      &\textbf{0.9956}  &\textbf{0.9791}  &\textbf{0.9436} & \textbf{0.9453}\\
  \midrule
 \multirow{10}{*}{LPIPS ($\downarrow$)}
 &0\%&Anchor      &0.1260     &0.1637  &0.0694     &0.0313    &0.0162 &0.1167 &0.0559     &0.0796 & 0.0823\\
 \cline{2-12}
 &      &\emph{base1}  &0.1548     &0.1816  &0.0883      &0.0591  & 0.0458 &0.1335  &0.0805  & 0.1102 & 0.1067\\  
 &5\%   &\emph{base2}  &\textbf{0.1487}     &0.1756  &0.0831     &0.0561  & \textbf{0.0364} &0.1352  &0.0719  & 0.1046 & 0.1015\\
 &      &LeHoPP &0.1496     &\textbf{0.1743}  &\textbf{0.0786}      &\textbf{0.0423}   &0.0536  & \textbf{0.1221}  &\textbf{0.0667}  & \textbf{0.0927} & \textbf{0.0975}\\
 \cline{2-12}
 &      &\emph{base1}  &0.1834     &0.2018  &0.1108      &0.0897      &0.0779  &0.1519 &0.1073  &0.1404 & 0.1329\\  
 &10\%   &\emph{base2}  &\textbf{0.1749}     &\textbf{0.1909} &0.1008      &0.0845     &\textbf{0.0606}  &0.1534 &0.0916  &0.1323  & 0.1236\\
 &      &LeHoPP &0.1853      &0.1946 &\textbf{0.0909}      &\textbf{0.0628}     &0.1077   &\textbf{0.1296}  &\textbf{0.0813}  &\textbf{0.1136}& \textbf{0.1207}\\
 \cline{2-12}
 &      &\emph{base1}  &0.2397     &0.2471      &0.1627      &0.1547     &0.1434     &0.1908  &0.1648  &0.2007 & 0.1878\\  
 &20\%   &\emph{base2}  &\textbf{0.2319}    &\textbf{0.2286}      &0.1446      &0.1448     &\textbf{0.1167}     &0.1910  &0.1388  &0.1920 & \textbf{0.1736}\\
 &      &LeHoPP &0.2781     &0.2544      &\textbf{0.1242}      &\textbf{0.1329}     &0.2517     &\textbf{0.1483}  &\textbf{0.1189}  &\textbf{0.1756} & 0.1855\\
  \bottomrule
\end{tabular}
\end{table*}

\subsection{Experimental results}

Results with different pruning percentages $\gamma$ of $5\%$, $10\%$, and $20\%$ are shown in Tab.~\ref{tab:results}. They represent the rendered view quality for the target video frames, averaged among all views of a sequence. We can observe that, on average, LeHoPP outperforms both \emph{base1} and \emph{base2} for all the computed metrics, for all three given values of $\gamma$. 
We further notice that, even though $\gamma$'s value of $20\%$ significantly reduces the number of source pixels in the rendering process, the performance is not very far from the anchor, where rendering is done on the source input without any pruning. Our method performs particularly well on sequences Carpark, Fan, Hall, Street, and Mirror. However, we observe slightly lower effectiveness in the case of sequences Painter, Frog, and Shaman. Some elements of suboptimality that justify this behavior (for the sake of computational efficiency) can be found, for example, in the inpainting approximation used for the importance computation, introduced in \eqref{eq:inp}, or in the one-shot nature of the pruning process itself (making multiple smaller steps would make the approximation in \eqref{eq:taylor} more precise) and, most importantly, not fine-tuning IBRNet to be more robust to the pruned pixels.

\subsection{Analysis and discussion}
\begin{figure}
\centering
  \includegraphics[trim={10 0 30 40},clip,width=0.9\columnwidth]{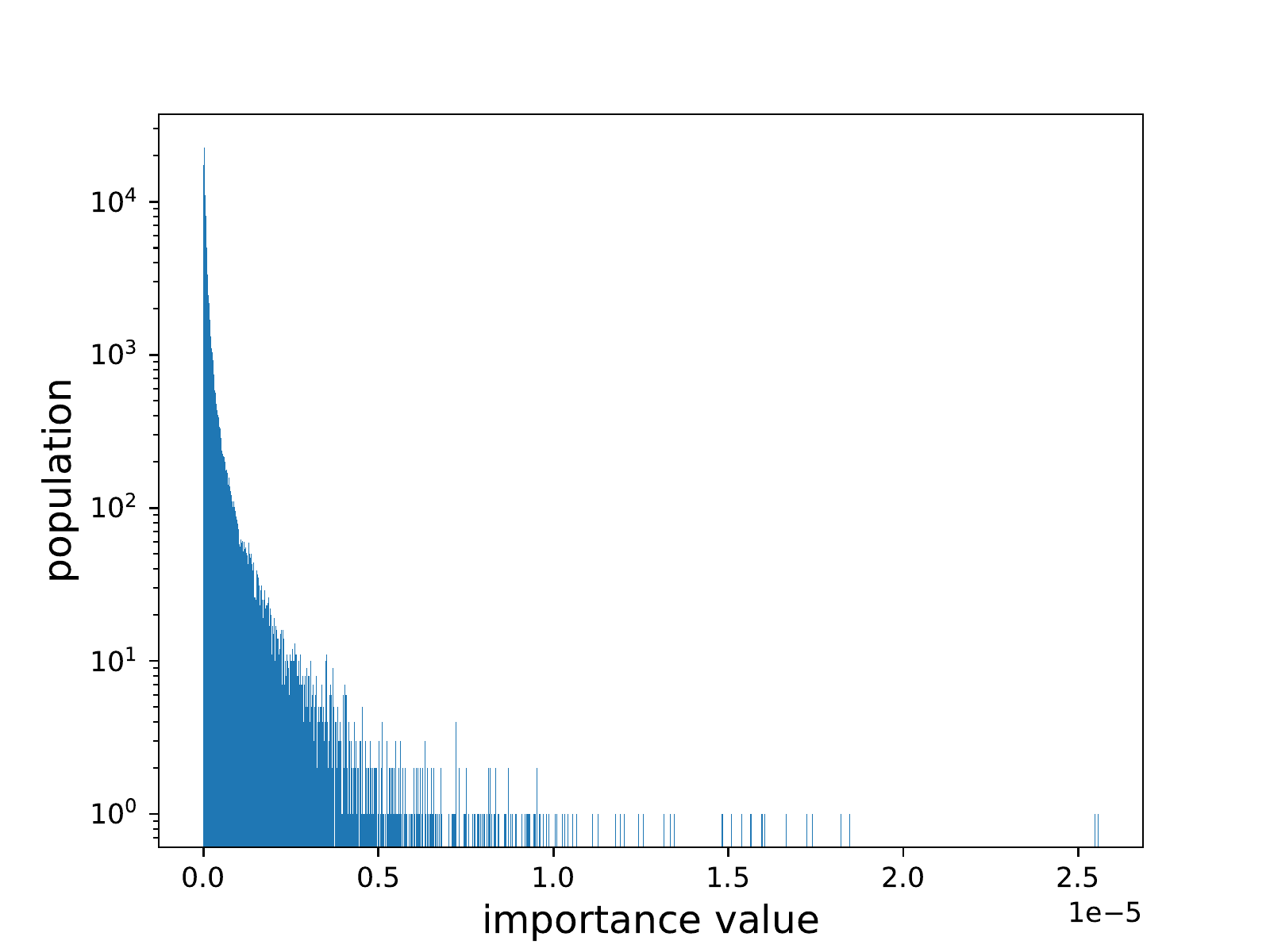}
\caption{Histogram of the importance values, \texttt{Street} video sequence, \texttt{view 1 at frame 0}.}
\label{fig:histogram}
\end{figure}

\begin{figure}[ht!]
    \centering 
\begin{subfigure}{0.25\textwidth}
  \includegraphics[trim={800 100 50 100},clip,width=0.95\linewidth]{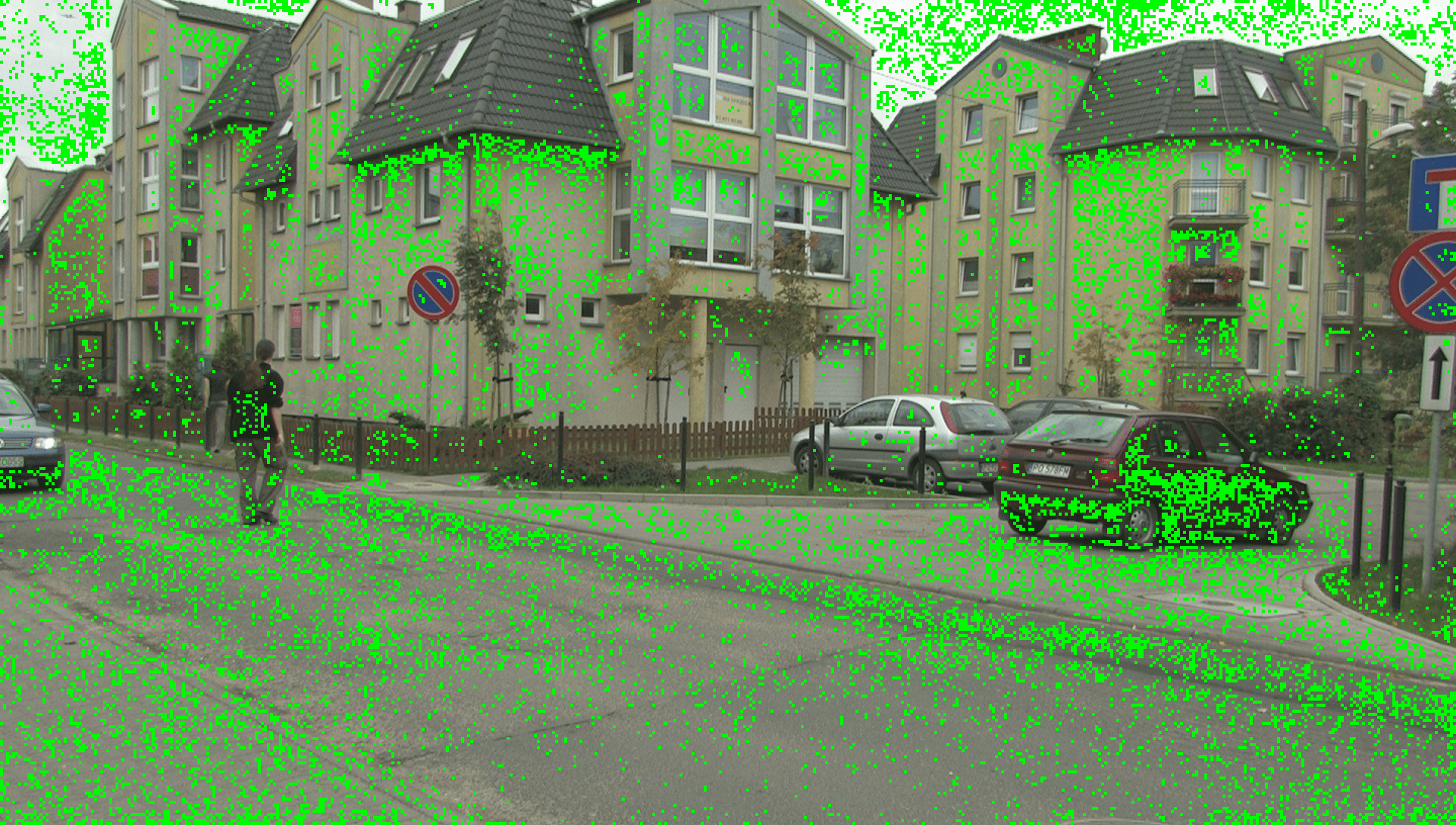}
  \caption{Pruned input \\($\gamma=20\%$)}
  \label{fig:1}
\end{subfigure}
\begin{subfigure}{0.25\textwidth}
\captionsetup{justification=centering}
  \includegraphics[trim={800 100 50 100},clip,width=0.95\linewidth]{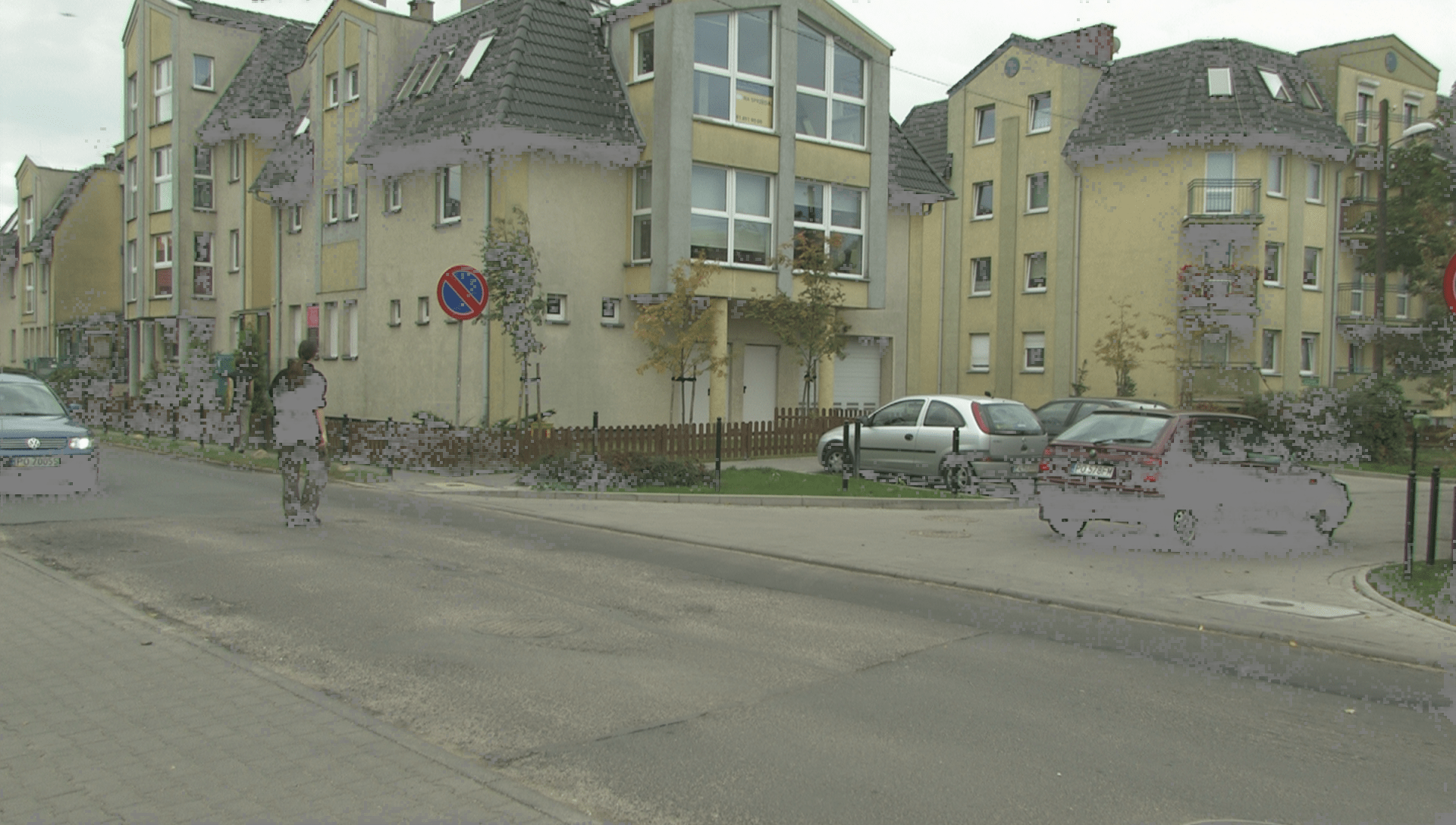}
  \caption{Output without input inpainting \\($21.35$ dB)}
  \label{fig:2}
\end{subfigure}

\begin{subfigure}{0.25\textwidth}
\captionsetup{justification=centering}
  \includegraphics[trim={800 100 50 100},clip,width=0.95\linewidth]{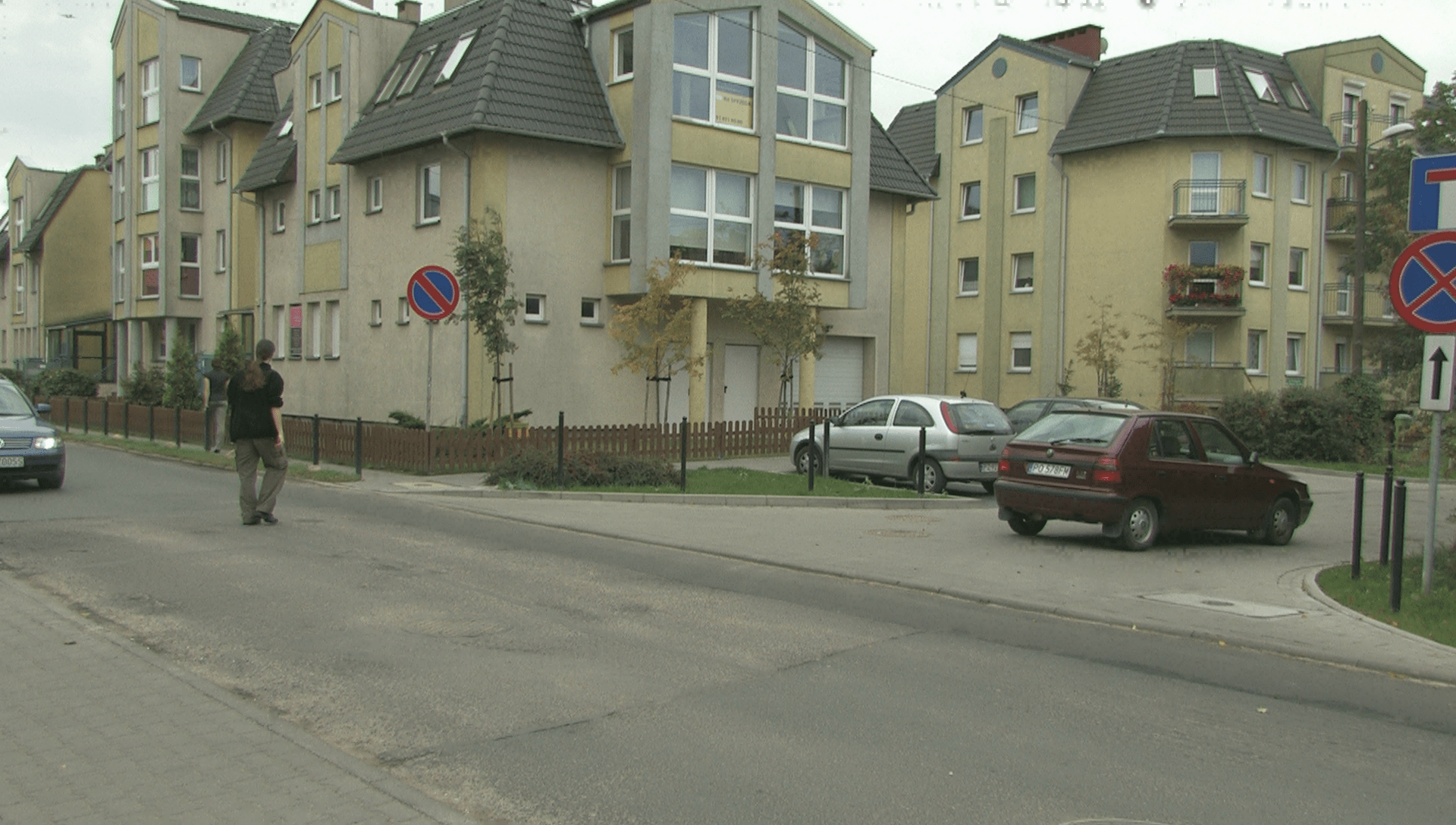}
  \caption{Pruned input with inpainting\\($\gamma=20\%$)}
  \label{fig:3}
\end{subfigure}
\begin{subfigure}{0.25\textwidth}
\captionsetup{justification=centering}
  \includegraphics[trim={800 100 50 100},clip,width=0.95\linewidth]{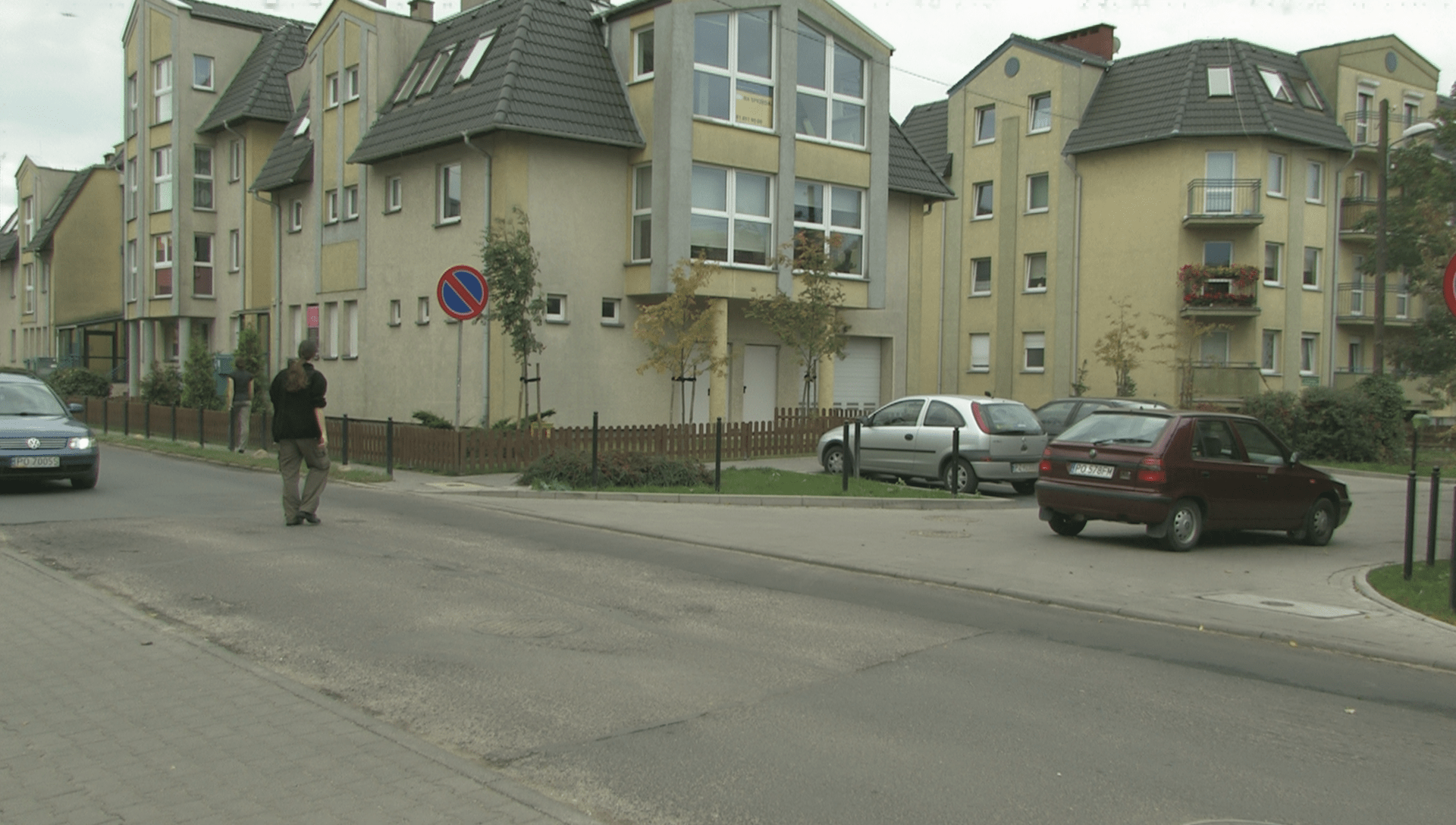}
  \caption{Output with input inpainting\\($40.73$ dB)}
  \label{fig:4}
\end{subfigure}
\caption{Qualitative comparison of a target image for a given input view: pruned pixels in (a) are in green for visualization. }
\label{fig:inpainting}
\end{figure}
The main contribution of this work, pixel importance computation, has proven to be a reliable metric for pixel pruning. Let us observe closely the contribution of the key elements of LeHoPP and how consistently it is better than the considered baselines, on a sample scene: we provide here an in-depth analysis using the Street sequence.

Fig.~\ref{fig:histogram} shows the distribution of the importance values $\mathcal{I}_i$: the majority of pixels (population) have a very small importance value, showing that it is, in principle, possible to prune a large number of pixels.

Replacing the pruned pixels with their inpainted counterparts before rendering helps the network to renderer the views with higher quality. Fig.~\ref{fig:inpainting} compares the quality of synthesized views of the Street sequence, one without (Fig.~\ref{fig:2}) and another with inpainting prior to rendering (Fig.~\ref{fig:4}).

\begin{figure}
\centering
  \includegraphics[trim={3 0 40 40},clip,width=\columnwidth]{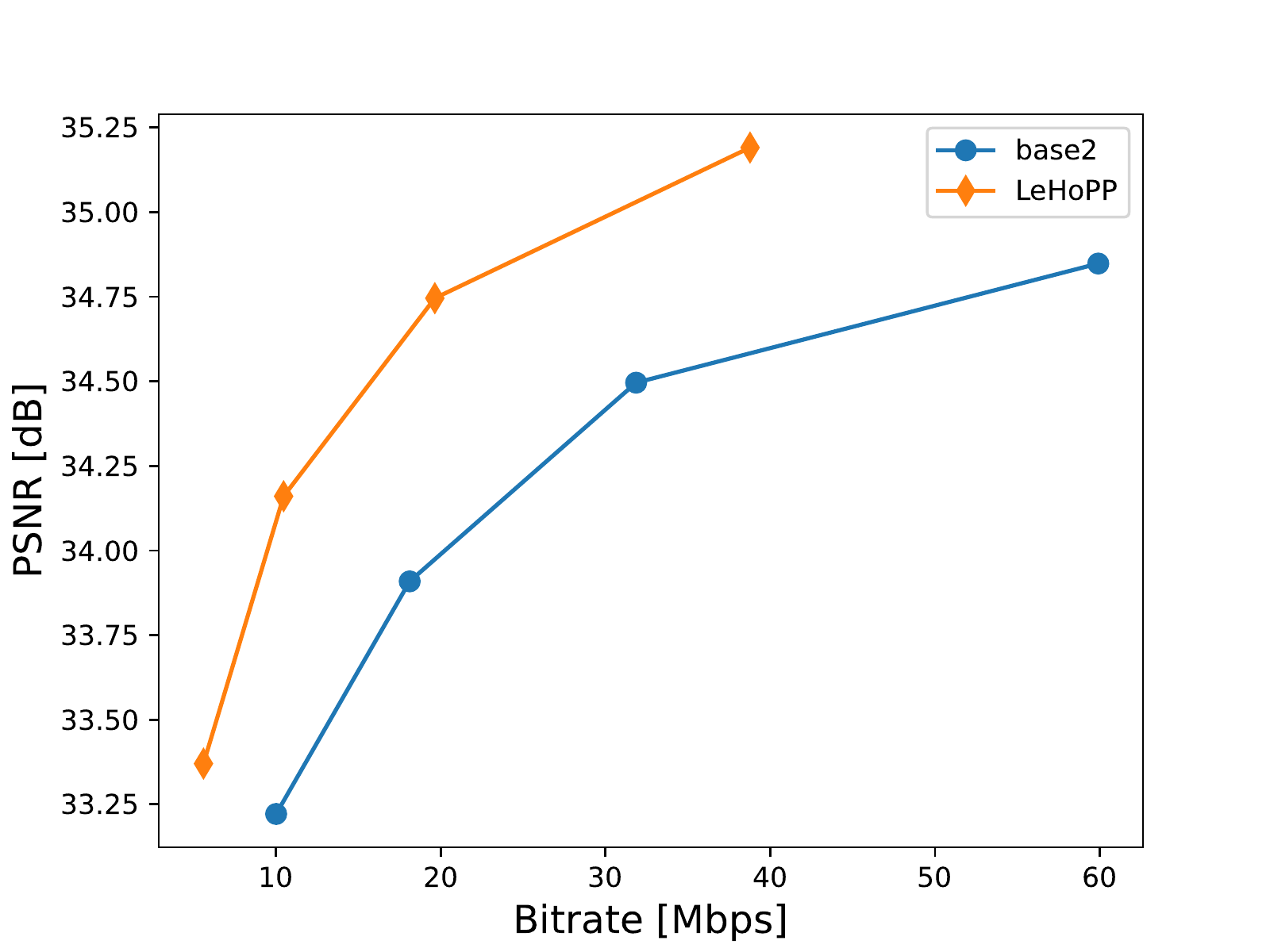}
\caption{RD performance of the LeHoPP method as compared to \emph{base2} on \texttt{Street} video sequence.}
\label{fig:RD-perf}
\end{figure}

Additionally, to measure the performance of LeHoPP compared to \emph{base2} in terms of data rate, we conduct a compression study. The setup of this study is HEVC simulcast, with $\text{QP}\!\in\!\{22,27,32,37\}$. First, all views of the Street sequence are pruned with $\gamma\!=\!10\%$; then they are compressed, and finally the decoded pruned views are given to the IBRNet renderer, without inpainting. LeHoPP yields $54.24\%$ BD-rate~\cite{BD-rate} against \emph{base2}, and performs consistently better in both quality and bitrate, for a given QP (Fig.~\ref{fig:RD-perf}).
 
Pixel rate in the MIV standard~\cite{pixel-rate} is limited to $32$ Mega pixels at $30$ fps. Sequences with high resolution and a large number of source views are barely fulfilling this requirement, if we consider solely the transmission of the source texture views without their corresponding depth maps. Therefore, pruning percentages between $5\%$ and $20\%$, tested in the scope of this work, are well-fitting to show the real possible use case scenario, where the amount of preserved pixels stays in the defined pixel rate boundaries. 

The inpainting approximation that we use for the computation of importance score is a proxy for inpainting, and a good simplification in the case when pruned areas are smaller. Therefore, results in Tab.~\ref{tab:results} show higher gains of LeHoPP compared to \emph{base1} and \emph{base2} on smaller $\gamma$ values. Furthermore, during importance computation, fewer views were used as a source input for rendering of Shaman, Painter, Frog, Mirror, and Fan, due to memory limitations and big backpropagation computational graph. This might also result in less precise importance values for these sequences. Moreover, for importance values, the method requires a few minutes of computation per frame, for a specific target view.

The TMIV pruning technique depends on both the depth maps and textures, and it is created to optimize for DIBR. Moreover, the MIV standard cannot transmit all the source views of a captured scene. Thus, it is incompatible with a neural renderer such as IBRNet, even though they are compatible in terms of data: the GA profile of MIV can transmit multi-view texture-only videos. Hence, there is a need to prune the textures for the IBRNet renderer for a potential transmission use case. The proposed system is not a standalone immersive video coding setup, but a first, necessary pre-processing stage of such a system. Our study demonstrates the method that learns which pixels are not essential and reduces the pixel rate while maintaining a high synthesis quality. 

\section{Conclusion}
\label{conclusion}
In this paper, we propose LeHoPP, a new approach that learns how to prune the pixels and can be utilized with neural image-based synthesizers in a multi-view setup without retraining the model. LeHoPP leverages the existing redundancy among different views of a scene and uses IBRNet to compute rendered view loss and its impact on each input pixel, thereby computing the pruning mask for each view, given the desired pruning percentage. Our method outperforms random block-based pruning baselines and shows good efficiency on the MPEG immersive video dataset without scene fine-tuning. We believe that our work encourages further development of neural renderers and broader adoption of standards such as MIV.
Future work could improve the pruning mask computation and integrate the pruning module into the training pipeline for better performance.

\section{Acknowledgement}
This work has been partially supported by the European Union under the Italian National Recovery and Resilience Plan (NRRP) of NextGenerationEU, partnership on ``Telecommunications of the Future'' (PE0000001 - program ``RESTART'').

\bibliographystyle{IEEEtran}
\bibliography{main}

\end{document}